%% file: Ankit_Pat_Thesis.tex
\documentclass[12pt,lot, lof]{puthesis}
\pdfoutput=1
\usepackage{subfigure}
\usepackage{hyperref,xspace}
\usepackage[pdftex]{graphicx}
\usepackage[cmex10]{amsmath}
\usepackage{algorithmic}
\usepackage[ruled,vlined]{algorithm2e}
\usepackage{array}
\usepackage{mdwmath}
\usepackage{amsmath}
\usepackage{mdwtab}
\usepackage{eqparbox}
\newcommand{\proquestmode}{}

\title{On Construction of a Class of Orthogonal Arrays}

\submitted{May 2012}  
\author{Ankit Pat}
\adviser{Professor Somesh Kumar}  
\department{Mathematics}

\usepackage{graphicx}

\usepackage{verbatim}

\usepackage{multirow}
\usepackage{longtable}

\usepackage{booktabs}

\ifdefined\printmode

\usepackage{url}

\else

\ifdefined\proquestmode

\usepackage{hyperref}
\hypersetup{bookmarksnumbered}

\makeatletter
\hypersetup{pdftitle=\@title,pdfauthor=\@author}
\makeatother

\else

\usepackage{hyperref}
\hypersetup{colorlinks,bookmarksnumbered}

\makeatletter
\hypersetup{pdftitle=\@title,pdfauthor=\@author}
\makeatother

\fi 
\fi 

\ifodd 0
\renewcommand{\maketitlepage}{}
\renewcommand*{\makecopyrightpage}{}
\renewcommand*{\makeabstract}{}
\renewcommand*{\makecertificate}{}

\else

\certificate{
\input{certificate}
}
\abstract{
\input{abstract}
}

\acknowledgements{
\input{acknowledgements}
}

\dedication{To my parents.}

\fi  

\begin{document}

\makefrontmatter


\include{chapter-intro}
\include{chapter-pastwork}

\include{chapter-algo}

\include{chapter-results}
\include{chapter-conclusion}
\ifdefined\phantomsection
    \phantomsection
  \else
  \fi
  \newpage
  \addcontentsline{toc}{chapter}{References}
  \begin{flushleft}
  \vspace*{0.25in}
  \Huge {\textbf{References}}
  \vspace*{0.1in}
  \end{flushleft}
  \begin{enumerate}
  \singlespacing
\item Bose R. C., \& Bush, K. A. (1952). Orthogonal arrays of strength two and three. \textit{Ann. Math. Statist.}, \textbf{23}, 508-524.\\
\item Bush, K. A. (1952). Orthogonal arrays of index unity. \textit{Ann. Math. Statist.}, \textbf{23}, 426-434.\\
\item Hamming, R. W. (1950). Error-detecting and error correcting codes. \textit{Bell Syst. Tech. J.}, \textbf{29}, 147-160.\\
\item Hedayat, A. , Sloane, N.J.A. and Stufken, J. (1999).Orthogonal Arrays. \textit{Springer-Verlag}, New York.\\
\item Rao, C. R. (1947). Factorial experiments derivable from combinatorial arrangements of arrays. \textit{J. Royal Statist. Soc. (Suppl.)}, \textbf{9}, 128-139.\\
\item Rao, C. R. (1949). On a class of arrangements. \textit{Proc. Edinburgh Math. Soc.}, \textbf{8}, 119-125.\\
\item Shrikhande, S.S. (1964). Generalised Hadamard matrices and orthogonal arrays of strength two. \textit{Canadian J. Math}, \textbf{16}, 736-740. \\
\item Sinha, K., Kumar, S., Sen Gupta, A. (2009). Construction of ternary orthogonal arrays by Kronecker sum. \textit{J. Stat. \& Appl.} \textbf{4(2-3)}, 475-478. \\
\item Sinha, K., Mathur, S. N. and Nigam, A. K. (1979). Kronecker sum of incomplete block designs. \textit{Utilitas Mathematica}, \textbf{16}, 157-164.\\
\item Sinha, K., Vellaisamy, P., Sinha, N. (2008). Kronecker sum of binary orthogonal arrays. \textit{Utilitas Mathematica,} \textbf{75}, 249-257.\\
\item Wang, J. C. and Wu, C. F. J. (1991). An approach to construction of asymmetrical orthogonal arrays. \textit{J. Amer. Statist. Assoc.}, \textbf{86}, 450-456. \\
\item Zhang, Y., Weiguo, L., Mao, S., Zheng, Z. (2006). A simple method for constructing orthogonal arrays by the Kronecker sum. \textit{J. Syst. Sci. Complexity}, \textbf{19}, 266-273.\\ 

\end{enumerate}
  \clearpage




\end{document}

%% file: certificate.tex
This is to certify that the thesis entitled \textit{On Construction of a Class of Orthogonal Arrays} submitted by \textbf{Mr. Ankit Pat} for the partial fulfillment of the requirements for the degree of Master of Science in \textbf{Mathematics and Computing} from the Department of Mathematics, Indian Institute of Technology, Kharagpur, is an authentic record of the work carried out under my supervision and guidance.
\\
\\
\vspace*{0.3in}
\begin{flushright}
Prof. Somesh Kumar\\
Department of Mathematics\\
Indian Institute of Technology\\
Kharagpur, India - 721302\\
\end{flushright}

%% file: abstract.tex
We propose a novel method for the construction of orthogonal arrays. The algorithm makes use of the Kronecker Product operator in association with unit column vectors to generate new orthogonal arrays from existing orthogonal arrays. The effectiveness of the proposed algorithm lies in the fact that it works well with any linear seed orthogonal array without imposing any constraints on the strength or the number of levels. The resulting orthogonal array has the same strength as the seed orthogonal array. We also discuss the proof of correctness of the algorithm. In the Results section we provide a list of new orthogonal arrays generated using this algorithm, that are currently not present in the libraries of orthogonal arrays.

%% file: acknowledgements.tex
I would like to express my profound and sincere gratitude to my supervisor, Professor Somesh Kumar, Department of Mathematics, Indian Institute of Technology, Kharagpur.  His constant guidance and constructive comments went a long way towards shaping the present thesis. I would also like to thank Professor Kishore Sinha, Department of Statistics \& Mathematics, Birsa Agricultural University, Ranchi, for sharing his expertise and providing his important support throughout this work.\\

I also wish to thank my friends for all the help and encouragement. Last but not the least, I would like to thank my parents for being there for me and supporting me throughout my life in all its aspects.
\\
\\
\vspace*{0.3 in}
\begin{flushright}
Ankit Pat\\
Department of Mathematics\\
Indian Institute of Technology\\
Kharagpur, India - 721302\\
\end{flushright}

%% file: chapter-intro.tex
\chapter{Introduction\label{ch:intro}}

Construction of orthogonal arrays (OAs) is an important problem in combinatorial design which holds great significance for design of experiments in statistical analysis. In the past, several construction methods for generating orthogonal arrays have been proposed and analyzed. Hedayat et al.  (1999) provide a comprehensive study of orthogonal arrays. The significance of the Kronecker Product and the Kronecker Sum operations in the context of generating orthogonal arrays is well established, as seen in the works of Shrikhande (1964), Wang and Wu (1991), Zhang, Weiguo, Mao and Zheng (2006) and Sinha, Vellaisamy and Sinha (2008). Sinha et al. (2009) used the Kronecker Sum operation on ternary orthogonal arrays and Balanced Incomplete Block Designs to construct new symmetrical ternary orthogonal arrays. In the current work we propose a novel construction approach for orthogonal arrays using unit column vectors and the Kronecker Product operations on existing orthogonal arrays.

\section{Preliminaries and Definitions}
\subsection{Balanced Arrays}
\par A balanced array, denoted by $BA(N,m,s,t)$ $ \{  \mu_{x_1,....,x_t}  \} $, is defined as an $N \times m$ matrix $B$ with elements belonging to the set $S = \{0,1,....,s-1\}$ of $s$ symbols, $m$ factors, $N$ runs and strength $t$ such that every $N \times t$ sub-matrix of $B$  contains the ordered row vector $(x_1,....,x_t)$, $ \mu_{x_1,....,x_t}$ times, where $\mu_{x_1,....,x_t}$ is invariant under any permutation of $x_1,....,x_t$. 

\subsection{Orthogonal Arrays}
\par An $N \times k$ array $A$ with entries from $S$ is said to be an orthogonal array with $s$ levels, strength $t \ ( 0 \leq t \leq k )$ and index $\lambda$ if every $N \times t$ sub-array of $A$ contains each $t$-tuple based on $S$ exactly $\lambda$ times as a row.
\par If $ \mu_{x_1,....,x_t} =\mu$(constant) $ \forall \  t-tuples \ (x_1,\hdots,x_t) \in S$ in a balanced array, then the balanced array becomes an orthogonal array with index $\mu$. 

\subsection{Kronecker Sum}
\par Sinha et al. (1979) defined the Kronecker sum of matrices $A$ (order $m \times n$) and $B$ (order $p \times q$) as $A\otimes B = A \otimes J + J \otimes B$ , where $\otimes$ denotes the usual Kronecker product and $J$ is a matrix with all its elements unity but is of dimension $p \times q$ in the first term and $m \times n$ in the second term. The Kronecker sum of binary orthogonal arrays was defined by Sinha et al. (2008).

\subsection{Galois Field} 
\par A Galois field is a field that contains finite number of elements and is denoted by $GF(s)$, where $s$ is its order. The order of a Galois field is the number of elements in the field and is of the form $p^n$, where $p$ is a prime number known as characteristic of the field and n is a positive integer. We shall denote the elements of the $GF(s)$ by $\{0, 1, 2,...., s-1 \}$ and the set of all $n$-tuples with entries from $GF(s)$ by $GF(s)^n$.

\subsection{Simple Orthogonal Arrays}
\par An orthogonal array is said to be simple if all its runs are distinct.

\subsection{Linear Orthogonal Arrays}
\par Let $s=p^n$, where $p$ is a prime and $n$ is a positive integer. Then, the orthogonal array $OA(N,k,s,t)$ with levels from $GF(s)$ is linear if it satisfies the following two conditions:
\begin{itemize}
\item it is simple 
\item when its rows are considered as $k$-tuples from the $GF(s)$, its $N$ runs form a vector space over $GF(s)$.
\end{itemize}

%% file: chapter-pastwork.tex
\chapter{Related Work\label{ch:pastwork}}

The problem of generating orthogonal arrays has been of interest to researchers since more than half a century. A number of algorithms have been proposed and constructions given. The orthogonal arrays that have been successfully constructed using such constructions are stored in libraries. In this chapter we shall discuss some important results in this area of research so as to build a basic framework before discussing the work presented in this thesis.
We shall discuss the theorems and constructions briefly without digressing by going into the details of the proofs. For detailed proofs the reader is directed to the references.
\section{The Rao-Hamming Construction}
As the name indicates, this construction was given by Rao (1947, 1949) and Hamming (1950), both of whom had found this algorithm independently. Rao had originally introduced a special case of the idea of orthogonal arrays in a rather implicit sense in his concepts of hypercube of strength $t$. His construction of hypercubes of strength $2$ find relevance and correspondence with orthogonal arrays $OA(s^n,(s^n-1)/(s-1),s,2)$, $s$ being a prime power. This brings us to the following theorem.

\textbf{Theorem 2.1:} If $s$ is a prime power then an $OA(s^n,(s^n-1)/(s-1),s,2)$ exists whenever $n \geq 2$.

\textbf{Construction:} Let us consider an $s^n \times n$ array whose rows are all possible $n$-tuples over $GF(s)$. Let, $C_1, C_2,.....,C_n$ be the columns of this array. The columns of the orthogonal array then consist of all the columns of the form 
\[ z_1 C_1 + z_2 C_2 + ..... + z_n C_n = [C_1, C_2,.....,C_n] z , \]
where $z=(z_1, z_2, ......, z_n)^T$ is an $n$-tuple from $GF(s)$, not all the $z_i$ are $0$, and the first $z_i$ is 1. Then, there are $(s^n-1)/(s-1)$ such columns.  
\section{Bush's Construction}
Bush's (1952) research on orthogonal arrays of index 1 is a well known and important result. The theorem proposed by Bush goes as follows.

\textbf{Theorem 2.2:} If $s \geq 2$ is a prime power then an $OA(s^t,s+1,s,t)$ of index unity exists whenever $s \geq t-1 \geq 0$.

The result stated in the above theorem can be improved for some values of $s$ and  $t$ as can be seen below.

\textbf{Theorem 2.3:} If $s=2^m, m \geq 1,$ and $t=3$ then there exists an $OA(s^3,s+2,s,t).$
 
The orthogonal arrays constructed in Theorems 2.2 and 2.3 are simple and linear.

\textbf{Theorem 2.4:} If $s$ is a prime power and a linear array $OA(s^t,k,s,t)$ exists, then there also exists a linear array $OA(s^{k-t},k,s,k-t).$
 
 
 

\section{Bose and Bush's Recursive Construction}
This construction was proposed by Bose and Bush (1952). It allows for the construction of orthogonal arrays of strength two with a large number of factors and possibly the the maximal number, provided that the number of symbols $s$ and the index $\lambda$ are powers of the same prime.
The theorem is stated as follows:

\textbf{Theorem 2.5:} Let $s = p^v$ and $\lambda = p^u$, where $p$ is a prime and $u$ and $v$ are integers with $u \geq 0, v \geq 1$. Let $d = \lfloor u/v \rfloor$. Then there exists an 
\[ OA(\lambda s^2, \lambda (s^{d+1}-1)/(s^d - s^{d-1}) + 1, s, 2)\]  
\section{Hadamard Matrices and Orthogonal Arrays}

A Hadamard matrix is a square matrix which takes only two symbols +1 and -1 as its entries such that for every two different rows there are matching entries in exactly half of the cases and non-matching entries in the remaining half. It can be easily noticed that Hadamard matrices are difference schemes with two symbols. Hadamard matrices and orthogonal arrays have close resemblances in their combinatorial properties. Hadamard matrices can be generated using recurrence relations. One such method is called Sylvester's method which goes as follows:

Let $H_i$ denote a Hadamard matrix of order $i$. Then,
\[ H_{1} = \begin{bmatrix} 1 \end{bmatrix}\]
\[ H_2 = \begin{bmatrix} 1 & 1 \\ 1 & -1 \\ \end{bmatrix} \]
\[ \vdots \]
\[ H_{2^k} = \begin{bmatrix} H_{2^{k-1}} & H_{2^{k-1}} \\ H_{2^{k-1}} & -H_{2^{k-1}} \\ \end{bmatrix} \]

An important result that illustrates the close connection between Hadamard matrices and orthogonal arrays is mentioned below.

\textbf{Theorem 2.6:} Orthogonal arrays $OA(4 \lambda, 4 \lambda -1, 2, 2)$ and $OA(8 \lambda, 4 \lambda, 2, 3)$ exist if and only if there exists a Hadamard matrix of order $4 \lambda$.


%% file: chapter-algo.tex
\chapter{A Novel Construction Algorithm\label{ch:algo}}

In this chapter, we propose a novel method for the  construction of orthogonal arrays using existing orthogonal arrays and unit column vectors with the help of the Kronecker Product operator. The method proposed herein serves towards forming orthogonal arrays of larger dimensions from orthogonal arrays of smaller dimensions. We primarily deal with linear orthogonal arrays, however as we will show later, non-linear orthogonal arrays can also be constructed using this approach.
We then investigate the correctness of the construction and other possible extensions to the algorithm.
\section{The Construction}

Let $A$ denote a linear seed $OA(N,k,s,t)$ with levels from $GF(s)$. The construction provides an approach to generate another $OA(N^{2}, k^{2}+2k, s, t)$.
Let $C = \{ c_{1}, c_{2}, \hdots, c_{k} \}$ denote the set of all the factors of $A$. Now let us define another zero column vector $c_{k+1}$. Let $U$ denote a unit column vector of dimensions $N \times 1$. See $Algorithm \ 1$ for the detailed construction.  

 \begin{algorithm}
\DontPrintSemicolon
\SetKwFunction{Construct}{Construct}
\SetKwFunction{return}{return}
\Begin{
$ i \leftarrow 1$ \\
$ j \leftarrow 1$ \\
$C = C \cup \{ c_{k+1} \}$\\
$ C^{\prime} = \phi $ \\

 \While{$i \leq k+1 \ $}
 {\While{$j \leq k+1 \ $}{
 	\If{$i \neq k+1$ or $j \neq k+1$}{  
	$c^{\prime} = c_{i} \otimes U + U \otimes c_{j}$ \\
	$c^{\prime} = c^{\prime} ( $mod$ \ s )  $ \\
	$C^{\prime} = C^{\prime} \cup \{ c^{\prime} \} $\\
	$ j \leftarrow j+1 $ \\
	}
 }
 $ i \leftarrow i+1 $ 
}
return $C^{\prime}$
}
\caption{The Construction Procedure on (A,C,U,s)}
\end{algorithm}

$Algorithm \ 1$ returns a set of column vectors $C^{\prime}$. Then, $C^{\prime}$ forms the set of all the factors of an $OA(N^{2}, k^{2}+2k, s, t)$. Let us denote this $OA(N^{2}, k^{2}+2k, s, t)$ as $B$.

\section{Correctness of the Algorithm}

Now let us discuss the proof of correctness of the construction proposed above. 
We shall state the result formally in the form of the following theorem.\\

\textbf{Theorem 3.1:} The existence of a linear orthogonal array $OA(N,k,s,t)$ with levels from $GF(s)$ implies the existence of an orthogonal array $OA(N^2, k^2 + 2k, s, t).$ \\

\textbf{Proof.} \ We proceed with the proof by first showing that the $OA(N^2, k^2 + 2k, s, t)$ generated from the seed linear orthogonal array $OA(N,k,s,t)$ is itself linear. Recall the denotations used in the previous section since they will be used again in this proof. \\
We can write $B=[B_1 | B_2 | \hdots | B_k | B_{k+1} | B_{k+2} ]$, where $B_i$ is defined as follows:
\[
  B_i = \left\{ 
  \begin{array}{l l}
    \{ c^{\prime}_{i} | c^{\prime}_{i} = ( c_{i} \otimes U + U \otimes c_{j} ) mod \ s, \ \forall j = 1 \ to \ k  \} & \quad \text{for $i=1$ \ to \ $k$}\\
    \{ c^{\prime}_{k+1} | c^{\prime}_{k+1} = ( U \otimes c_{j} ) mod \ s , \ \forall j = 1 \ to \ k  \} & \quad \text{if $i= k + 1$}\\
    \{ c^{\prime}_{k+2} | c^{\prime}_{k+2} = ( c_{j} \otimes U ) mod \ s , \ \forall j = 1 \ to \ k  \} & \quad \text{if $i= k + 2$}\\
  \end{array} \right.
\]


Let $R_j=[b_{1}^{j}|b_{2}^{j}| \hdots |b_{k+2}^{j}]$ be the $j^{th}$ row of $B$, such that $b_{1}^{j},b_{2}^{j},\hdots,b_{k+2}^{j}$ are the $j^{th}$ rows of $B_1, B_2 , \hdots , B_{k+2}$ respectively.
Then due to the definition of the construction, it can be seen that the following results hold:
\begin{itemize}
\item for $i=1 \ to \ k$, $b_{i}=r+\beta_{i} U_{1 \times k}$ 
\item $b_{k+1} = r$ 
\item $b_{k+2}=[\beta_{1}|\beta_{2}| \hdots |\beta_{k}]$
\end{itemize}
where $r$ and $[\beta_{1}|\beta_{2}| \hdots |\beta_{k}]$ are rows of $A$ and $U_{1 \times k}$ is a unit row vector. 

Thus we can find rows $r_1$, $r_2$, $[\beta_{1}|\beta_{2}| \hdots |\beta_{k}]$ and $[\beta_{1}^{\prime}|\beta_{2}^{\prime}| \hdots |\beta_{k}^{\prime}]$ of $A$ which give the the $i^{th}$ and $j^{th}$ rows of $B$ as follows:
\[R_i=[b_{1}^{i}|b_{2}^{i}| \hdots |b_{k+2}^{i}]= [ r_1+\beta_{1}U_{1 \times k}|r_1+\beta_{2}U_{1 \times k}| \hdots |r_1+\beta_{k}U_{1 \times k}|r_1|\beta_{1}|\beta_{2}| \hdots |\beta_{k}]\]
 \[R_j=[b_{1}^{j}|b_{2}^{j}| \hdots |b_{k+2}^{j}]= [ r_2+\beta_{1}^{\prime}U_{1 \times k}|r_2+\beta_{2}^{\prime}U_{1 \times k}| \hdots |r_2+\beta_{k}^{\prime}U_{1 \times k}|r_2|\beta_{1}^{\prime}|\beta_{2}^{\prime}| \hdots |\beta_{k}^{\prime}]\]
 
 A linear combination of the above two rows using scalars $h_1$ and $h_2$ from the $GF(s)$ will result in the following: \\
 $h_1 R_i + h_2 R_j = [ h_1 r_1+ h_2 r_2+(h_1 \beta_1+h_2 \beta_{1}^{\prime})U_{1 \times k}|h_1 r_1+ h_2 r_2+(h_1 \beta_2+h_2 \beta_{2}^{\prime})U_{1 \times k}| \hdots |h_1 r_1+ h_2 r_2+(h_1 \beta_k+ h_2\beta_{k}^{\prime})U_{1 \times k}|h_1 r_1+ h_2 r_2|h_1 \beta_1+h_2 \beta_{1}^{\prime}|h_1 \beta_2+h_2 \beta_{2}^{\prime}| \hdots |h_1 \beta_k+h_2 \beta_{k}^{\prime}]$\\
 
 Now, since A is known to be linear, and $r_1$ and $r_2$ are rows of $A$, $h_1 r_1 + h_2 r_2$ is also a row of A. Similarly, $[h_1 \beta_1+h_2 \beta_{1}^{\prime} | h_1 \beta_2+h_2 \beta_{2}^{\prime} |  \hdots  | h_1 \beta_k+h_2 \beta_{k}^{\prime}]$ is a row of A. Hence, by definition of the construction and its properties discussed earlier in this proof, $h_1 R_i + h_2 R_j$ is a row of $B$. It can also be easily noted that one row of $B$ is a zero vector of $1 \times (k^2+2k)$ size. Hence, it can be concluded that the rows of $B$ form a linear subspace of $GF(s)^{k}$.
  
 So, let $N^2=s^n$ and let $G$ be the generator matrix for $B$ of dimensions $n \times (k^2+2k)$ such that the rows of $B$ consist of all $k$-tuples $\eta G$, where $\eta \in GF(s)^n$. Now, suppose there exist $t$ columns of $B$ that are linearly dependent over $GF(s)$. Then, due to the nature of the construction, there exist $x$ $(1 \leq x \leq t)$ columns in $A$ that are linearly dependent, which is a contradiction. Hence, every selection of $t$ columns of $B$ are linearly independent over $GF(s)$. So, let us choose $t$ columns of $B$ and let $G_1$ be the corresponding submatrix of generator matrix $G$. Then, the columns of $G_1$ will be linearly independent. Also, the number of times a $t$-tuple $\tau$ is present as a row in these $t$ columns of $B$ is determined by and is equal to the number of $\eta$ such that $\eta G_1 = \tau$.  
 
 Now, since $G_1$ has rank $t$, the number of such $\eta$ is $s^{(n-t)}$, for all $\tau$. Hence, $B$ is an orthogonal array of strength $t$. This concludes the proof. \hfill 

\section{Binary Orthogonal Arrays}

In this section we propose the following lemma, wherein we show that the linearity condition in Theorem 1 can be dropped for binary seed orthogonal arrays (i.e. OAs of strength 2). 

\textbf{Lemma 3.1:}  The existence of an $OA(N,k,s,2)$ implies the existence of an $OA(N^{2}, k^{2}+2k, s, 2)$. 

\textbf{Proof.} Let $A$ be $OA(N,k,s,2)$ as described before and let $c^{\prime}_{1}$ and $c^{\prime}_{2}$ be any two distinct factors of $B$. Also, let us assume $c^{\prime}_{1}$ and  $c^{\prime}_{2}$ were generated from the factors $c_{i_{1}}$, $c_{j_{1}}$, $c_{i_{2}}$, $c_{j_{2}}$ of $A$ using the following equations: 
\[ c^{\prime}_{1} = ( c_{i_{1}} \otimes U + U \otimes c_{j_{1}} ) mod \ s \]
\[ c^{\prime}_{2} = ( c_{i_{2}} \otimes U + U \otimes c_{j_{2}} ) mod \ s \]
Let $c^{\prime}_{1}$ and  $c^{\prime}_{2}$ be denoted by $[X_{1} : X_{2} :  \hdots  : X_{N}]^{\prime}$ and $[Y_{1} : Y_{2} :  \hdots  : Y_{N}]^{\prime}$, where the dimensions of $X_{i}$ and $Y_{i}, \  \forall i = 1 \ to \  n$, is $N \times 1$. 

In order to show the validity of the algorithm we split the analysis into four cases as follows.\\

$Case \ 1: \ $ $c_{j_{1}} \neq c_{j_{2}}$ such that $c_{j_{1}} , c_{j_{2}} \neq c_{k+1}$ \\

We know that $X_{i} = (c_{j_{1}} + p_{i} U)mod \ s$ and $Y_{i} = (c_{j_{2}} + q_{i} U)mod \ s$, where $p_{i}, q_{i} \in \{ 0, 1, 2,  \hdots , s-1 \},  \  $ for all $i = 1 \ to \  N$. Also $c_{j_{1}} \neq c_{j_{2}}$ and $c_{j_{1}}, c_{j_{2}} \in C$ implies that $c_{j_{1}}$ and $c_{j_{2}}$ form the factors of an $OA(N,2,s,2)$. Since, for all $i = 1 \ to \  N, \  X_{i}$ and $Y_{i}$ are just cyclic permutations of symbols of  $c_{j_{1}}$ and $c_{j_{2}}$, the sets $\{ X_{i}, Y_{i} \}$ form the factors of Orthogonal Arrays with parameters $(N,2,s,2)$. Therefore, $Z = [c^{\prime}_{1} : c^{\prime}_{2}]$ being a vertical juxtaposition of the matrices $[X_{i} : Y_{i}]$ for $i = 1 \ to \  N$, is an $OA(N^2,2,s,2)$. Hence, $c^{\prime}_{1}$ and  $c^{\prime}_{2}$ form the factors of an $OA(N^2,2,s,2)$. \\

$Case \ 2: \ $  $c_{j_{1}} = c_{j_{2}}$ such that $ c_{j_{1}} , c_{j_{2}} \neq c_{k+1}$ \\

$X_{i} = (c_{j} + p_{i} U)mod \ s$ and $Y_{i} = (c_{j} + q_{i} U)mod \ s$, where $c_{j_{1}} = c_{j_{2}} = c_{j}$ (say) and $p_{i}, q_{i} \in \{ 0, 1, 2,  \hdots , s-1 \},  \  $ for all $i = 1 \ to \  N$. Since we know that $c_{i_{1}} \neq c_{i_{2}}$, and also the fact that $c_{i_{1}}$ and $c_{i_{2}}$ together form the factors of an $OA(N,2,s,2)$, we therefore can conclude that all possible 2-tuples on $ \{0,1,2, \hdots ,s-1 \} $ belong to the set $ \{ (p_i,q_i) : i = 1 \ to \  N \} $ and also that each tuple occurs exactly $N/s$  times. Hence, $Z = [c^{\prime}_{1} : c^{\prime}_{2}]$ which is a vertical juxtaposition of the matrices $[X_{i} : Y_{i}]$ for $i = 1 \ to \  N$, is an $OA(N^2,2,s,2)$and thus $c^{\prime}_{1}$ and  $c^{\prime}_{2}$ form the factors of an $OA(N^2,2,s,2)$.\\

$Case \ 3: \ $  $c_{j_{1}} =  c_{k+1}$ and $c_{j_{2}} \neq c_{k+1}$ \\

$c^{\prime}_{1}$ is a juxtaposition of $\alpha U, \alpha \in \{0,1,2, \hdots ,s-1\}$ with each $\alpha U$ appearing $N/s$ times.  $c^{\prime}_{2}$ is a juxtaposition of $c_{j_{2}} + \beta U, \beta \in \{0,1,2, \hdots ,s-1\}$ where each  $c_{j_{2}} + \beta U$ has symbols from $\{0,1,2, \hdots ,s-1\}$ an equal number of times (since $A$ is orthogonal of order $1$). Hence $c^{\prime}_{1}$ and $c^{\prime}_{2}$ contain all possible 2-tuples over $ \{0,1,2, \hdots ,s-1 \} $ an equal number of times.\\

$Case \ 4: \ $  $c_{j_{1}} =  c_{j_{2}} = c_{k+1}$ \\

$c_{j_{1}} =  c_{j_{2}}$ implies $c_{i_{1}} \neq  c_{i_{2}} $. Also since $c_{i_{1}} $ and $ c_{i_{2}} $ contain all possible 2-tuples over $\{0,1,2,\hdots,s-1\}$ (since $A$ is orthogonal of order $2$), $c^{\prime}_{1}$ and $c^{\prime}_{2}$ contain all possible 2-tuples over $ \{0,1,2,\hdots,s-1 \} $ equal number of times.\\

Since $c^{\prime}_{1}$ and  $c^{\prime}_{2}$ were chosen arbitrarily, $C^{\prime}$ forms the set of all the factors of an $OA(N^{2}, k^{2}+2k, s, 2)$. Hence, the construction given in $Algorithm \ 1$ gives an $OA(N^2,k^2+2k,s,2)$ for every existing $OA(N,k,s,2)$.\\

It can be noticed that the linearity constraint is not necessary for proving that the method gives an $OA(N^{2}, k^{2}+2k, s, 2)$. Hence the above lemma is proved.\\

\section{Generating Non-Linear Orthogonal Arrays}

Although we originally proposed the method for generating linear orthogonal arrays, it is possible to generate non-linear orthogonal arrays using the same construction. For accomplishing this we must take a linear orthogonal array to begin with, choose any one of its columns and cyclically permute its symbols. Then clearly the resulting orthogonal array will loose its linearity property. However, even then the construction will work well for this kind of seed orthogonal array and generate a non-linear orthogonal array. We discuss further about this approach and its correctness.   

Given a linear orthogonal array $A(N,k,s,t)$, we randomly choose a column $c_i$ of which we shall cyclically permute the symbols. After undergoing a cyclic permutation, the new column $c_i^{\prime}$ can be represented as follows:
\[ c_i^{\prime} = c_i + \alpha U\]
where $\alpha$ is a non-zero element of $GF(s)$ and $U$ is a unit column vector of dimensions $N \times 1$.

Then the columns of the orthogonal array $B$ generated using $c_i^{\prime}$ from the seed array $A$ may be of the form:

\[ c^{\prime} = ((c_i + \alpha U) \otimes U + U \otimes c_j ) mod s = (c_i \otimes U + U \otimes c_j  + \alpha U \otimes U) mod s\]

Irrespective of whether $c_j$ is a zero column vector or is equal or not equal to $c_i^{\prime}$, we always get a symbol-wise cyclic permutation of the column of $B$ we would have generated if $A$ were linear. Similarly, the result holds for 
\[ c^{\prime} = ((c_i + \alpha U) \otimes U + U \otimes c_j ) mod s \]

Since the the columns of the array we have obtained is proved to be equal to or a cyclic permutation of symbols of the columns of the linear orthogonal array which we would otherwise have generated, the array thus produced is orthogonal. Further, since it does not contain a zero vector for a row, it is not linear. Hence, we can successfully generate non-linear orthogonal arrays using the aforementioned construction.

%% file: chapter-results.tex
\chapter{Results\label{ch:results}}
The significance of a new construction lies in its ability to generate new orthogonal arrays which can be contributed to the existing libraries. Hence, it is important to show that the construction proposed in this work is actually capable of generating new orthogonal arrays. Thereby, in this chapter we shall discuss about the new orthogonal arrays that can be constructed using the proposed method. Comprehensive libraries of orthogonal arrays can be found in Hedayat $et \ al.$ (1999) or on the following websites:
\begin{center}
 \textit{http://www2.research.att.com/~njas/oadir/} \\
 \textit{http://support.sas.com/techsup/technote/ts723.html} \\
\end{center}

\section{List of New Contributions}

In this section we shall list out tables of orthogonal arrays which are generated using the proposed construction but are not found in any of the above mentioned libraries.
 
\begin{table}[ht]
\caption{Orthogonal Arrays with Two Levels and Strength 2} 
\centering  
\begin{tabular}{c c c} 
\hline\hline                        
Sl. No. & Seed OA & Generated OA \\ [0.5ex] 
\hline                  
! & OA(8,5,2,2) & OA(64,35,2,2) \\
2 & OA(8,7,2,2) & OA(64,63,2,2) \\
3 & OA(12,11,2,2) & OA(144,143,2,2) \\
4 & OA(16,15,2,2) & OA(256,255,2,2) \\
5 & OA(20,19,2,2) & OA(400,399,2,2) \\ [1ex]      

\hline 
\end{tabular}
\label{table:nonlin} 
\end{table}

\begin{table}[ht]
\caption{Orthogonal Arrays with Two Levels and Strength 3} 
\centering  
\begin{tabular}{c c c} 
\hline\hline                        
Sl. No. & Seed OA & Generated OA \\ [0.5ex] 
\hline                  
1 & OA(24,12,2,3) & OA(576,168,2,3) \\
2 & OA(32,16,2,3) & OA(1024,288,2,3) \\
3 & OA(40,20,2,3) & OA(1600,440,2,3) \\
4 & OA(48,24,2,3) & OA(2304,624,2,3) \\
5 & OA(56,28,2,3) & OA(3136,840,2,3) \\ 
6 & OA(64,32,2,3) & OA(4096,1088,2,3) \\
7 & OA(72,36,2,3) & OA(5184,1368,2,3) \\ [1ex]      

\hline 
\end{tabular}
\label{table:nonlin} 
\end{table}

\begin{table}[ht]
\caption{Orthogonal Arrays with Two Levels and Strength $>$  3} 
\centering  
\begin{tabular}{c c c} 
\hline\hline                        
Sl. No. & Seed OA & Generated OA \\ [0.5ex] 
\hline                  
1 & OA(80,6,2,4) & OA(6400,48,2,4) \\
2 & OA(128,9,2,5) & OA(16384,99,2,5) \\
3 & OA(64,7,2,6) & OA(4096,63,2,6) \\ [1ex]      

\hline 
\end{tabular}
\label{table:nonlin} 
\end{table}

\begin{table}[ht]
\caption{Orthogonal Arrays with Three Levels and Strength $\geq$  2} 
\centering  
\begin{tabular}{c c c} 
\hline\hline                        
Sl. No. & Seed OA & Generated OA \\ [0.5ex] 
\hline                  
1 & OA(27,13,3,2) & OA(729,195,3,2) \\
2 & OA(81,40,3,2) & OA(6561,1680,3,2) \\
3 & OA(54,5,3,3) & OA(2196,35,3,3) \\ [1ex]      

\hline 
\end{tabular}
\label{table:nonlin} 
\end{table}

\begin{table}[ht]
\caption{Orthogonal Arrays with More Than Three Levels} 
\centering  
\begin{tabular}{c c c} 
\hline\hline                        
Sl. No. & Seed OA & Generated OA \\ [0.5ex] 
\hline                  
1 & OA(16,5,4,2) & OA(256,35,4,2) \\
2 & OA(64,21,4,2) & OA(4096,483,4,2) \\
3 & OA(64,6,4,3) & OA(4096,48,4,3) \\
4 & OA(25,6,5,2) & OA(625,48,5,2) \\
5 & OA(49,8,7,2) & OA(2401,80,7,2) \\ 
6 & OA(64,9,8,2) & OA(4096,99,8,2) \\
7 & OA(81,10,9,2) & OA(6561,120,9,2) \\
8 & OA(121,12,11,2) & OA($11^4$,168,11,2) \\
9 & OA(169,14,13,2) & OA($13^4$,224,13,2) \\
10 & OA(256,17,16,2) & OA($2^{16}$,288,16,2) \\ 
11 & OA(289,18,17,2) & OA($17^4$,360,17,2) \\ [1ex]      

\hline 
\end{tabular}
\label{table:nonlin} 
\end{table}

%% file: chapter-conclusion.tex
\chapter{Conclusion and Future Work\label{ch:conclusion}}

In this work, we presented a novel construction algorithm for generating orthogonal arrays from seed linear orthogonal arrays. The proposed method works well for seed linear orthogonal arrays of all strengths and levels. We also discussed the proof of correctness of the construction and also the possibility of generating non-linear orthogonal arrays using the same construction. The results show that the proposed construction is indeed capable of contributing to the existing libraries of orthogonal arrays. Lists of new orthogonal arrays generated using this method have been provided in the Results section.\\

Extensive experimental observations suggest that the proposed construction works well even if the required linearity condition is dropped. This gives us reason to conjecture that the construction would work well for any seed orthogonal array. We have already proved this result for seed orthogonal arrays of strength 2.  Our future work will be primarily directed towards proving the correctness of the construction for non-linear seed orthogonal arrays of any strength.